# How to image single isolated atoms by using coherent low-energy electrons

Tatiana Latychevskaia

*Paul Scherrer Institute, Forschungsstrasse 111, 5232 Villigen, Switzerland*

*Physics Department, University of Zurich, Winterthurerstrasse 190, 8057 Zurich, Switzerland*

tatiana@physik.uzh.ch*, tatiana.latychevskaia@psi.ch*

## Abstract

Coherent low-energy electrons have been demonstrated as a practical tool for imaging individual macromolecules and two-dimensional (2D) crystals. Low-energy electrons exhibit unique properties: low radiation damage to biological molecules and high sensitivity to the local potentials. In this study, we outline the conditions at which single isolated charge-free atoms can be imaged by low-energy electron holography. A single atom produces an interference pattern consisting of concentric fringes of finite diameter and of very weak intensity. The diffraction angle $\theta$, determined as the first minimum of the concentric rings interference pattern, exhibits similar dependency on the source-to-sample distance zs as $\sin\theta \sim 0.3/z_s^{1/2}$ for electrons of different energies (50, 100 and 200 eV) and scattered off different elements (Li, C, and Cs). The results are compared to the recently reported experimental holograms of alkali atoms intercalated into bilayer graphene and adsorbed on top of graphene.

## 1. Introduction

Low-energy electron holography (20 – 250 eV) (Fig. 1a,b) has been demonstrated for imaging individual macromolecules and two-dimensional crystals [1]. Single biological macromolecules, like DNA, membranes, tobacco mosaic virus, ferritin, and others, have been successfully imaged by low-energy electrons [2-13]. In some of these experiments, biological macromolecules were exposed to low-energy electrons even for hours without detectable structural changes [14, 15], while the same sample would be immediately destroyed in a conventional transmission electron microscope (TEM). However, the resolution in the reconstructions obtained from low-energy electron holograms is just below 1 nm [11, 16], and therefore, the effect of radiation damage on the structures at the atomic level cannot be characterized. Theoretical studies of the bond breakage in biological molecules when exposed to low-energy electrons are provided in refs. [17-20].

Presently, it is assumed that the main reasons why the observed resolution in low-energy electron holography is lower than the theoretical resolution are the vibrations of the sample or the electron source. The resolution of an object reconstructed from its hologram is determined by the visibility of the fine interference fringes in the hologram. A shift by 1 Å of either the object or the electron source (Fig. 1b) will result in a shift of the hologram by 100 μm. Assuming a $10^6$ magnification factor of the setup and a detector pixel size of 10 μm, it gives a shift by 10 pixels. Such rando fine vibrations of the source or the sample will lead to hologram interference pattern being randomly shifted, and as a result, the total recorded hologram will exhibit smeared out fine interference fringes. Since the resolution of the reconstructed object is determined by the visibility (contrast) of the fine interference fringes, the resolution will be reduced. In addition, the sample itself can undergo fast structural changes that also cause blurring of the interference pattern. The use of cryo-temperatures [21] or an advanced anti-vibrational environment [13] did not improve the obtained resolution. Recently, it was discussed that atomic resolution in images of three-dimensional objects, such as, for example, single proteins, cannot in principle be obtained due to the strong diffraction of low-energy electrons and as a result of strong out-of-focus signal [22].

Livadaru et al. suggested that it should be possible to obtain images of atomically resolved structures by low-energy electron holography [23]. Though, in their simulations, Livadaru et al. used somewhat unrealistic parameters: the source-to-sample distance was only 10 nm and the source-to-detector distance only 1 cm, while in typical low-energy electron microscopes these distances amount to tens of nanometers and 5 – 20 cm, respectively [1, 11, 13, 24, 25]. The lowest electron energy that was reported in a low-energy electron holography experiment was 18 eV, where the corresponding source-to-sample distance was evaluated by reconstructing the shown holograms and amounts to about 57 nm [26].

Another unique property of low-energy electrons is their strong interaction with matter [27] and high sensitivity to the local potentials in the sample. Charged adsorbates with a sub-elementary charge exhibit a 10 times higher intensity in low-energy electron holograms than when they would be imaged by electrons of conventional energies of 60 – 300 keV [28, 29]. This property of the low-energy electrons makes low-energy electron holography a unique tool for probing the structural and electronic properties of two-dimensional crystals [26, 27, 30].

Single individual adatoms and molecules on graphene can be imaged in a conventional TEM [31]. Individual atoms deposited onto graphene represent a perfect test object to study the limit of the resolution in low-energy electron holography. Previously, it was observed that charged adsorbates create distinct interference patterns in the holograms – bright spots with concentric fringes [28, 29]. In a different experiment, the time evolution of single alkali atoms intercalating into a graphene bilayer was recorded by low-energy electron holography [32], where the objects observed in holograms were interpreted as clusters of atoms. However, the question of imaging single atoms that are charge-free has not yet been investigated. No quantitative study was performed to show what type of interference patterns would be created by individual atoms and whether these interference patterns can be detected in an experiment.

In this study we consider the simplest test object: graphene with charge-free adsorbed atoms on top, and we outline the conditions at which single isolated atoms can be observed in low-energy electron holograms.

## 2. Low-energy electron holography principle

Coherent imaging with low-energy electrons (20 – 250 eV, Fig. 1a) is conventionally realized in the Gabor or in-line type holography scheme, which does not employ any lenses, and thus the recorded holograms are free from the associated aberrations (Fig. 1b) [1, 9, 33, 34]. The coherent electrons are extracted by field emission from a sharp metal tip, typically tungsten. The sharpness of the tip defines the spatial coherence of the electron wave [35]. The source-to-sample distance can be varied and typically ranges from tens of nanometers to a few microns. The energy of the probing electrons depends on the extracting voltage and decreases when the source-to-sample distance decreases. The sample is illuminated by a divergent spherical wave, and thus, it is imaged in a projection mode at the magnification given by $M = z_d/z_s$, where $z_s$ is the source-to-sample distance, and $z_d$ is the source-to-detector distance. For example, for $z_s$ = 100 nm and $z_d$ = 100 mm, the magnification $M = 10^6$ and nanometer-sized objects can be imaged. The sample typically is an object that is placed on a supporting substance. The support is required to be transparent, which ensures that a large part of the probing wave does not interact with the sample and provides the reference wave [36, 37]. Graphene has been demonstrated as an ideal support for samples studied by low-energy electron holography: it is conductive and single-atom thick, therefore, causing minimal absorption of the probing wave [38-42]. It should be noted that even three layers of graphene already cause complete absorption of the low-energy electrons [16, 22], and thus, graphene is not just an ideal but probably the only possible support for objects to be imaged by low-energy electron holography. The hologram is formed by the interference between the wavefront scattered by the object and the not scattered wave and is recorded by a distant detector. The recorded holograms are then reconstructed numerically, and the structure of the object is recovered [43].

The wavelength of low-energy electrons, calculated as $\lambda = \dfrac{hc}{\sqrt{eV(2m_0c^2 + eU)}}$ is in the range between 0.78 – 2.24 Å for electron energies of 20 – 250 eV (Fig. 1a), where $h$ is Planck's constant, $c$ is the speed of light, $eU$ is the electron energy in $eV$, $e$ is the elementary charge, and $m_0$ is the electron mass. This range of wavelengths, in principle, allows for atomic-resolution imaging, for example, the interatomic distance in graphene is 1.42 Å. The lateral resolution of a low-energy holography setup evaluated by the Abbe criterion is given by $R = \lambda/(2NA)$ where NA is the numerical aperture of the setup NA≈$s_d/(2z_d)$, $s_d$ is the diameter of the detector. For the source-to-detector distance of $z_d$ = 0.18 m and the detector diameter 75 mm, the lateral resolution amounts to $\lambda/(2NA)$ = 1.9 – 5.49 Å for electron energies of 20 – 250 eV.

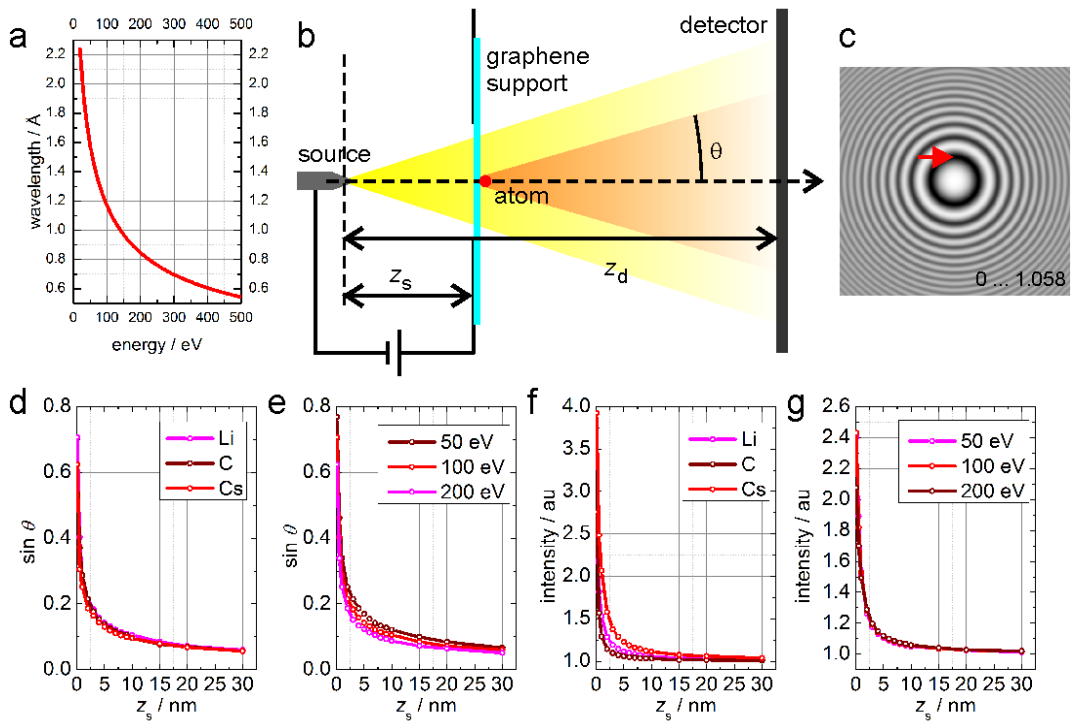

Fig. 1 Low-energy electron holography of single atoms. (a) Wavelength of the probing low-energy electrons as a function of their energy. (b) Experimental arrangement – a single atom on top of graphene, and (c) a corresponding hologram: simulated hologram of single Li atom imaged by 100 eV electrons at the source-to-sample distance $z_s$ = 10 nm, the first minimum (indicated by the red arrow) corresponds to the diffraction angle of $\sin\theta \approx 0.1$. (d) – (e) Diffraction angle ($\sin\theta$) as a function of the source-to-sample distance $z_s$ for Li, C, and Cs atoms probed with electrons of 100 eV energy (d), and for Li atoms probed with electrons of 50, 100 and 200 eV energy (e). (f) – (g) Intensity of the main maximum in the hologram (normalized by division with background, where the background is the hologram of graphene without the atom) as a function of the source-

to-sample distance $z_s$ for Li, C and Cs atoms probed with electrons of 100 eV energy (f), and for Li atoms probed with electrons of 50, 100, and 200 eV energy (g).

## 3. Holograms of single atoms, simulations

When imaging a single isolated atom supported by graphene, the following considerations can be made. Based on properties of far-field diffraction, one might expect that the smaller the object, the higher the diffraction angle. This, however, is not the case when the object is probed with a spherical wave. Here, a magnified image of the object is produced, similar to a projection imaging scheme. In addition, for coherent probing waves, not just a magnified image but an interference pattern is obtained at the detector plane. This interference pattern retains its distribution but scales up as the source-to-detector distance increases while the source-to-sample distance remains unchanged. On the other hand, this interference pattern will not just re-scale but change its distribution when the source-to-sample distance changes. At a very large source-to-sample distance, the distribution of the probing wave approaches that of a plane wave, and the resulting interference pattern will turn into a so-called Fresnel coherent diffraction pattern, where the central part is a hologram and the outer part (that does not contain the reference wave) exhibits the distribution of a diffraction pattern [44, 45]. An analytical solution to the optical integral describing this regime is not known. In this study, we apply numerical integration to obtain the distributions of the holograms of single atoms.

The simulations of holograms of single atoms deposited onto graphene were done as follows. Here, and in the rest of the text, we consider charge-free adsorbates. The transmission function of a single atom was modelled as $\exp\left[i\sigma v_z(x,y)\right]$, where $\sigma = \dfrac{2\pi m e \lambda}{h^2}$ is the interaction parameter, $m$ is the relativistic mass of the electron, $e$ is the elementary charge, $\lambda$ is the wavelength of the electrons, $h$ is the Planck constant, and $v_z(x,y)$ is the projected potential that was modelled using the parameters provided in ref. [46]. The presence of graphene support only reduces the background intensity by a constant factor and therefore is neglected. The test atom was placed at the center of the probing region for simplicity, taking into account that the distribution of the corresponding hologram should be invariant to the lateral position of the atom (in paraxial approximation, the wavefront propagation can be described as a convolution with a propagator function [43], and a lateral shift of the object causes a linear ramp phase term in the Fourier space that turns into a lateral shift of the object's hologram in the detector plane). The pixel size in the sample plane was set to 0.2 Å for correct sampling of the sample's projected potential. For such a small pixel size, the propagating function in the Fourier space can be only correctly sampled when the source-to-sample distance does not exceed 10 nm [43]. At larger source-to-sample distances, the propagating functions need to be sampled with tens of thousands of pixels to ensure correct sampling. However, in experiments, the source-to-sample distance typically ranges from 20 to 2000 nm. Thus, it is presently not possible to simulate a low-energy electron hologram of a single isolated atom at realistic experimental parameters due to the sampling issues. To solve this issue, the following approach is proposed and demonstrated in this study: the holograms are simulated at short source-to-sample distances using the correct sampling, and the parameters of the distributions of the obtained holograms are extrapolated for larger source-to-sample distances. In particular, two parameters are evaluated in the simulated holograms: the angle at which the first minimum occurs (indicated in Fig. 1c) which we call "diffraction angle" $\theta$, and the maxima of the intensity. The size of this distribution, and namely the position of the first minima, strongly depends on the source-to-sample distance, as discussed in detail below.

Holograms of single atoms were calculated for Li, C, and Cs atoms probed with 50 – 200 eV energy electrons at the source-to-sample distances 0.1 – 30 nm, Fig. 1d–g. The source-to-detector distance was kept the same in all simulations: $z_d$ = 1 m. The simulated holograms had different sizes, with the sidelength of a simulated hologram given by $s_d = z_d\, s_s/z_s$, where the sidelength in the sample plane was $s_s$ = 20nm in all the simulations. For each hologram, the position of the first minimum in the detector plane, $r_d$ (in meters), was evaluated from the hologram distribution, and the sine of the diffraction angle theta was evaluated from the known $z_d$ and $r_d$.

An example of a simulated hologram of a single isolated Li atom imaged with 100 eV electrons at a source-to-sample distance of 10 nm is shown in Fig. 1c. The hologram exhibits concentric fringes, and the first minimum is defined as the diffraction angle. The results in Fig. 1d – e show that the diffraction angles of electrons of different energies and scattered off different elements exhibit similar dependency on the source-to-sample distance. This can be explained by the fact that atomic potentials are described by functions that all show similar behaviours with the amplitude decay within 1 Å. From the simulated holograms, it was evaluated that the diffraction angle shows a dependency on the source-to-sample distance $z_s$, which can be described as a universal dependency $\sin\vartheta = 0.3/\sqrt{z_s}$, where $z_s$ is in nm; this function was obtained by fitting the data for Li atoms imaged with 100 eV electrons, and the other atoms/energies follow similar dependencies.

Once the function that describes the diffraction angle as a function of source-to-sample distance is obtained, it can be used to extrapolate the appearance of the hologram at larger, realistic source-to-sample distances. For example, for $z_s$ = 1650 nm, it gives $\sin\vartheta = 7.4\cdot10^{-4}$, which for the source-to-detector distance of 0.18 m gives the diameter of the main maxima of 2.66 mm. Thus, at these experimental parameters, a signature of a single atom in the hologram would be a concentric rings interference pattern with the diameter of the central disk of 2.66 mm.

The maximum of the intensity of the holograms of single charge-free atoms, shown in Fig. 1f – g, also exhibits similar dependency on the source-to-sample distance for different atoms and different energy, rapidly decreasing to the level of the background (one). From the simulated holograms, it was evaluated that the intensity maximum shows a dependency on the source-to-sample distance $z_s$ which can be described as $I_{max} = 0.3 z_s^{-0.8}$, where $z_s$ is in nm; this function was obtained by fitting the data for C atoms imaged with 100 eV electrons, and the other atoms/energies follow similar dependencies. From this dependency, it can be evaluated that, for example, for $z_s$ = 1650 nm, it gives $I_{max}$ = 8·10$^{-4}$, which is very weak contrast when compared to the background intensity of one. In order to distinguish this signal from the background, a very high signal-to-noise ratio in the hologram would be required.

## 4. Experimental holograms of intercalated atoms

Low-energy electron holograms of adsorbates on graphene were previously reported in refs. [28, 32]. Single atoms on graphene often exhibit a small charge and therefore are easily recognized in the low-energy electron holograms. Recently, low-energy electron holograms of alkali atoms (K, Li, Cs) intercalated into bilayer graphene were reported by Lorenzo et al. [32], an example of the hologram is shown in Fig. 2a. The hexagonal lattice visible in the lower bright half of the holograms shown in Fig. 2a is due to the fiber optic of the detector system. The corresponding reconstructions of the sample obtained by using the algorithms from ref. [43] are shown in Fig. 2b and c.

In the experiment, alkali atoms were spray-deposited in situ, and the holograms of the sample were recorded as a function of time. At the end of the experiment, the bilayer graphene region was fully covered by alkali atoms, while the region of the single-layer graphene was covered by a few adsorbates, which indicated that the adsorption on the single layer is energetically less favourable than the intercalation into the bilayer. In these holograms, concentric ring interference patterns of various intensities were observed; some are labelled in Fig. 2a. All these interference patterns were previously interpreted as holograms of clusters.

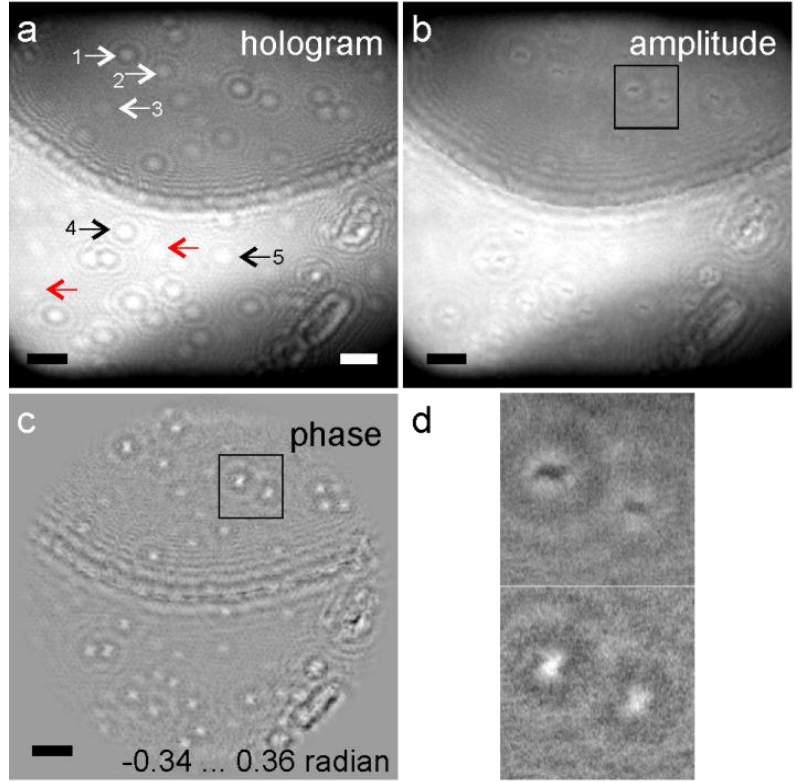

Fig. 2 Low-energy electron hologram of Li atoms intercalated into bilayer graphene and its reconstruction. (a) Experimental hologram, as previously reported in [32], acquired with 80 eV electrons; source-to-sample distance 1.65 µm, source-to-detector distance 180 mm. The scalebar on the left corresponds to 50 nm in the sample plane. The scalebar on the right corresponds to 5 mm in the hologram plane. (b) and (c) amplitude and phase distributions reconstructed from the hologram. Scalebars in (b) and (c) are 50 nm. (d) Magnified region from amplitude and phase reconstruction.

The holograms were acquired at the source-to-sample distance $z_s$ = 1650 nm, and the source-to-detector distance of 0.18 m, where the source-to-sample distance was determined by obtaining a sharp in-focus reconstruction of the edge of the graphene top layer. Some interference patterns that exhibit very weak contrast have the diameter of the main maximum of about 3.0 mm, which compares well with the expected 2.66 mm diameter obtained at these parameters in the simulated example above. Thus, these interference patterns can be created by diffraction at single Li atoms, examples of such interference patterns are indicated by the red arrows in Fig. 2a. In the supporting movies for the paper by Lorenzo et al., fast moving of such weak-intensity interference patterns (due to single atoms) and their merging with interference patterns of stronger intensity (due to clusters) can be observed during the Li atoms deposition [32].

In addition, it was checked whether the imaged object exhibits any absorption – by simply calculating the total intensity in the object region and comparing this intensity to the intensity of the background. In the absence of absorption by the object, the two values of intensity should be equal [22]. For the selected 5 spots in the hologram shown in Fig. 2a, the ratios of averaged intensity of the spot's interference pattern to the averaged intensity around the spot and the corresponding variance (indicated in parenthesis) were estimated: (1) 1.035 (0.154), (2) 1.002 (0.133), (3) 1.007 (0.121), (4) 1.002 (0.069) and (5) 1.000 (0.059). One important conclusion can be made from these numbers: all these evaluated ratios are equal to one or slightly larger, implying that the imaged objects do not exhibit any absorption; the values that exceed one can be explained by the error due to the noise.

Figure 2d shows a magnified region where two adsorbates (clusters of atoms) can be seen. Each adsorbate exhibit a non-symmetrical, elongated in one direction, bow-tie-like shape in the reconstructed amplitude and phase distributions. The reconstructed shape is elongated along orthogonal directions in the amplitude and phase distributions, which is an indication that neither distribution show the true shape of the object. Moreover, these reconstructed shapes appear to be elongated in various directions depending on the lateral position of the adsorbate in the sample, as it can be seen in Fig. 2b–d. In addition, an in-focus image of the adsorbates could not be obtained at any z-distance. All these distortions can be explained by the presence of local inhomogeneous electric or magnetic potentials [47].

## Conclusions

In conclusion, single isolated atoms deposited on graphene can, in principle, be detected by low-energy electron holography. A single atom can be recognized by its interference pattern consisting of concentric fringes of finite diameter and of very weak intensity. The experimental challenge is distinguishing this weak increase of intensity from that of the background, which requires a high signal-to-noise ratio of the recorded holograms. In the simulations, holograms of single atoms can be correctly simulated only for short source-to-sample distances of 1 – 10 nm. The parameters of the holograms obtained at larger, realistic source-to-sample distances can be estimated by extrapolating the distribution of the holograms to larger source-to-sample distances. Examples of holograms of single Li, C, and Cs atoms are provided, and the same simulations can be extended to other atoms.

# Acknowledgements

The authors acknowledge support by the Swiss National Foundation research grants 200021_197107 and 10003659.